# Bistable Circuit Behaviour as a 2 level (stable/metastable) potential energy system

Bosco Leung, Yon Chiet Ng, Safieddin Safavi-Naeini
ECE Department, University of Waterloo, Waterloo, Ontario, Canada

*Abstract*—
A novel model of analyzing Schmitt trigger as 2 level (metastable/stable) energy system is presented. The energy of the 2 level system is identified. In an MOS implementation, this arises from the electrostatic potential energy of the electrons on the gate to source capacitance of the cross coupled MOS transistors. The electrostatic energy, upon co-ordinate transformation, (incorporating the nonlinear MOS device behaviour and cross coupling), yields an expression for the 2 level system energy. The 2 level system, described under the transformed co-ordinate, is simulated. Simulation on the dynamics of the system shows a metastsable state and two stable states, corresponding to the a maximum(metastable) energy and a minimum(stable) energy, as predicted by the energy expression in the model. This result agrees with earlier attempt based on nonlinear dynamic system approach, whereby the potential is interpreted from the point of view of the gradient of vector field. In addition, the model gives physical design insight on how the change in circuit parameters (such as $g_m$, R), affects the energy characteristics, hence leads to effective design of such 2 level system.

## I. INTRODUCTION

Bistable circuit such as latch or Schmitt trigger is an important component in many electronic circuits such as astable multivibrators/relaxation oscillator, monostable multivibrators, comparators, sense amplifier/flip flop/cross coupled inverter pair [1]-[6].

When applying bistable circuit to systems where metastability is of consideration, as for example, being sense amplifier in memory, being cross coupled inverter pair for entropy source in True Random Number Generator (TRNG), it turns out basing the analysis on energy, rather than voltage and current of the circuit, is more fruitful. [1] illustrates qualitatively the bistable system such as Schmitt trigger to a physical analog as ball rolling from top to bottom of a hill as shown in Figure 1. The system exhibits metastability when the ball is resting at the top i.e. highest potential energy. A small disturbance on the ball causes the ball to roll down the hill and settles at the stable states, the lowest potential energy. A first attempt on analyzing this metastability is given in [7]. [7] interprets the potential from the point of view of the gradient of vector field. This is more a nonlinear dynamic system approach[8].

In this paper, we attempt to give physical meaning to the potential energy, as electrostatic energy, analogous to potential energy of the ball on top of a hill. This is used to show that, the metastability condition, corresponding to the highest potential energy, through regeneration, will have the system's energy lowered to the minimum energy, and settle to a stable condition.

Note, the application of variational principle, or minimizing of energy has also been recently applied in the circuit community for other applications[10].

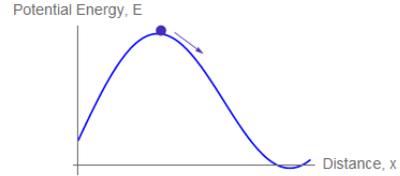

Figure 1 Potential Energy, E vs Distance, x. It illustrates the particle (round ball) always move "downhill" and settles to a position with the least potential energy along its trajectory.

## II. SCHMITT TRIGGER ( IN GROUND CAPACITOR BASED RELAXATION OSCILLATOR): PREVIOUS MODEL

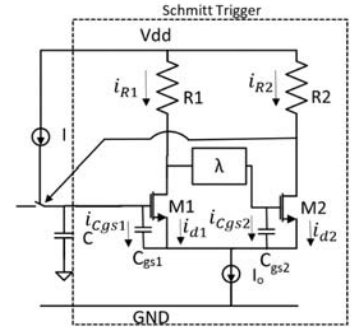

Figure 2 Ground capacitor relaxation oscillator model

Figure 2 shows the application of a Schmitt trigger in the design of a relaxation oscillator [7]. The Schmitt trigger governs the regeneration process in the switching of states for a relaxation oscillator. For simplification, $R_1 = R_2 = R$, $k_{n1} = k_{n2} = k_n$. Previously, [7] ignores $C_{gs2}$ and encapsulates the parasitic effects of $C_{gs1}$ as $\varepsilon = \frac{RC_{gs1}}{I_0}$ in (1). During the regeneration process Schmitt trigger is approximated by a first order differential equation in normalized current, $z = \frac{2i_{d1}-I_0}{I_0}$ as shown in (1).

$$\frac{dz}{dt} = \frac{1}{\varepsilon}(\sqrt{2}((1-g_mR)z + \frac{1}{8}z^3)) \qquad (1)$$

Then, [7] assumes (1) describes a first order system with a potential function [9], (denoted in [7] as E) such that

$$\frac{dz}{dt} = \frac{dE}{dz} \qquad (2)$$

By substituting (2) into (1), and integrating with respect to z, we have a first order ODE in E, with z, rather t, as variable.

$$E(z) = \frac{\sqrt{2}}{\varepsilon}\left(\frac{1}{2}(1-g_m R)z^2 + \frac{1}{32}z^4\right) \quad (3)$$

[7] identifies the first order ODE, $\frac{dE}{dz}$ as a gradient system and interprets 'E' as 'potential', in the sense of being the gradient of a vector field, which guides the equation of motion, but does not offer any physical meaning.

## II. NEW MODEL (SCHMITT TRIGGER AS A 2 LEVEL BISTABLE SYSTEM): DYNAMIC

In this paper, to model the Schmitt trigger as a 2 level system, both $C_{gs1}$ and $C_{gs2}$ are included, as shown in Figure 2. At the metastable state, $i_{d1} = i_{d2} = \frac{I_0}{2}$. $R_1 = R_2 = R$, $k_{n1} = k_{n2} = k_n$, and $\lambda = 1$.

The circuit is now described by two (rather than one) coupled first order differential equations, with both $C_{gs1}$ and $C_{gs2}$ are included, which is not done in [7]. By applying KCL at the drain nodes of both $M_1$ and $M_2$, (4), (5) and (6) are obtained:

$$i_{C_{gs1}} = i_{R2} - i_{d2} \quad (4)$$

$$i_{C_{gs2}} = i_{R1} - i_{d1} \quad (5)$$

$$i_{R1} + i_{R2} = I_0 \quad (6)$$

Next applying KVL around the loop of resistors, $R_1$, $R_2$, and transistors, $M_1$, $M_2$ we have:

$$i_{R1} - i_{R2} = \frac{v_{gs2} - v_{gs1}}{R} \quad (7)$$

By symmetrizing around the common mode signal, $\frac{I_0}{2}$, we obtain $i_{R1}$ and $i_{R2}$:

$$i_{R1} = \frac{I_0}{2} + \frac{v_{gs2} - v_{gs1}}{2R} \quad (8)$$

$$i_{R1} = \frac{I_0}{2} - \frac{v_{gs2} - v_{gs1}}{2R} \quad (9)$$

The capacitor current, $i_{C_{gs1}}, i_{C_{gs2}}$ are related to voltage, $v_{gs1}, v_{gs2}$ as shown below:

$$i_{C_{gs1}} = C_{gs1}\frac{dv_{gs1}}{dt}, i_{C_{gs2}} = C_{gs2}\frac{dv_{gs2}}{dt} \quad (10)$$

Transistor device equations using long channel approximation (ie. square law) give

$$v_{gs1} = \sqrt{\frac{2i_{d1}}{k_n}}, V_{gs2} = \sqrt{\frac{2i_{d2}}{k_n}} \quad (11)$$

The current flowing through the transistors, $i_{d1}$ and $i_{d2}$ can be related to the capacitor voltage, $V_{gs1}$ and $V_{gs2}$:

$$\frac{di_{d1}}{dt} = k_n\left(\sqrt{\frac{2i_{d1}}{k_n}} - v_t\right)\frac{dv_{gs1}}{dt},$$

$$\frac{di_{d2}}{dt} = k_n\left(\sqrt{\frac{2i_{d2}}{k_n}} - v_t\right)\frac{dv_{gs2}}{dt} \quad (12)$$

By substituting (8)-(12) into (4) and (5), two coupled differential equations are obtained as shown in (13) and (14).

$$C_{gs1}\frac{di_{d1}}{dt} = \frac{1}{k_n\left(\sqrt{\frac{2i_{d1}}{k_n}} - v_t\right)}\left(\frac{I_0}{2} + \frac{1}{2R}\left(\sqrt{\frac{2i_{d2}}{k_n}} - \sqrt{\frac{2i_{d1}}{k_n}}\right) - i_{d2}\right) \quad (13)$$

$$C_{gs2}\frac{di_{d2}}{dt} = \frac{1}{k_n\left(\sqrt{\frac{2i_{d2}}{k_n}} - v_t\right)}\left(\frac{I_0}{2} - \frac{1}{2R}\left(\sqrt{\frac{2i_{d2}}{k_n}} - \sqrt{\frac{2i_{d1}}{k_n}}\right) - i_{d1}\right) \quad (14)$$

Then, the phase portrait of an example of bistable system described in (13) and (14) is plotted in Figure 3. It shows that the Schmitt trigger has 1 metastable state at $\left\{\frac{i_{d1}}{I_0}, \frac{i_{d2}}{I_0}\right\} = \{0.5, 0.5\}$ and 2 stable states at $\left\{\frac{i_{d1}}{I_0}, \frac{i_{d2}}{I_0}\right\} = \{0, 1\}$ and $\left\{\frac{i_{d1}}{I_0}, \frac{i_{d2}}{I_0}\right\} = \{1, 0\}$. Hence it behaves a 2 level system. It is also shown that the dynamic of the system is symmetrical at $i_{d1} = i_{d2}$.

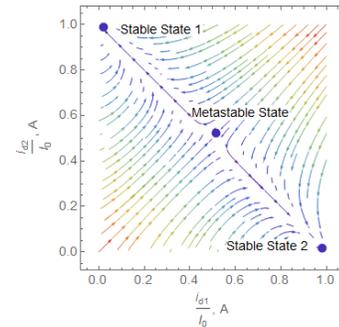

Figure 3 Phase portrait of (13), (14) using normalized Currents, $\frac{i_{d2}}{I_0}, \frac{i_{d1}}{I_0}$

Figure 4 shows the dynamic of the system going down from either side of the metastable state. As shown in Figure 4, the system trajectory either moves toward the stable state, $\left\{\frac{i_{d1}}{I_0}, \frac{i_{d2}}{I_0}\right\} = \{0, 1\}$ or moves toward the stable state, $\left\{\frac{i_{d1}}{I_0}, \frac{i_{d2}}{I_0}\right\} = \{1, 0\}$. The time evolution of the normalized currents, $\frac{i_{d1}}{I_0}, \frac{i_{d2}}{I_0}$ for the case of Figure 4a is shown in Figure 5.

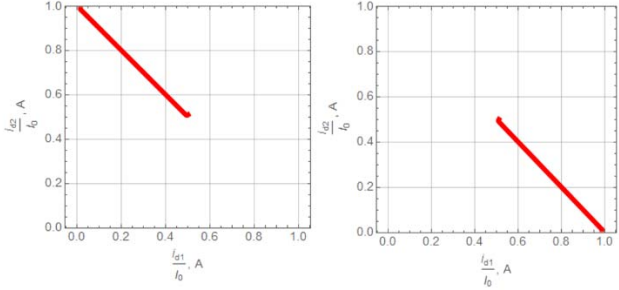

Figure 4 Phase curve of normalized Currents, $\frac{i_{d2}}{I_0}$ vs $\frac{i_{d1}}{I_0}$ (a) Red line is the trajectory of the Schmitt Trigger when the system dynamic moves from metastable state to stable state 1, (b) Red line is the trajectory of the Schmitt Trigger when the system dynamic moves from metastable state to state 2.

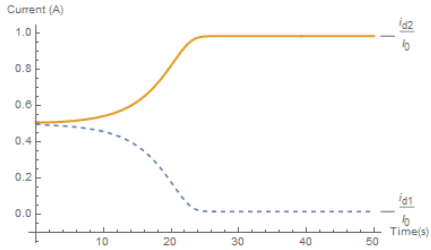

Figure 5 Time evolution of Current, $\frac{i_{d1}}{I_0}$ & $\frac{i_{d2}}{I_0}$ vs time, t

### III. NEW MODEL (SCHMITT TRIGGER AS A 2 LEVEL BISTABLE SYSTEM): POTENTIAL ENERGY

To obtain the potential energy, we push the development of the dynamics even further. We transform the co-ordinate, such that in this transformed co-ordinate, the resulting ODE has a similar form as that of some familiar physical system (such as forced spring/pendulum), which
a) Behaves as a 2 level system
b) have well known expression for potential energy.

By comparing the ODE describing the bistable Schmitt trigger in this transformed co-ordinate to that of the physical system, proper mapping can be established, which can be carried over to the energy expression.

To obtain the ODE in transformed co-ordinate, (14) is subtracted from (13), and (15) is obtained. Then we set $C_{gs1} = C_{gs2} = C_{gs}$ and replace both $k_n\left(\sqrt{\frac{2i_{d1}}{k_n}} - v_t\right)$ and $k_n\left(\sqrt{\frac{2i_{d2}}{k_n}} - v_t\right)$ by the same term $k_n\left(\sqrt{\frac{2i_d}{k_n}} - v_t\right)$ ($i_d$ is the average of $i_{d1}$ and $i_{d2}$, and so this term represents an 'average' of both $k_n\left(\sqrt{\frac{2i_{d1}}{k_n}} - v_t\right)$ and $k_n\left(\sqrt{\frac{2i_{d2}}{k_n}} - v_t\right)$ ) and we have:

$$\frac{C_{gs}}{k_n\left(\sqrt{\frac{2i_d}{k_n}} - v_t\right)}\frac{d(i_{d1}-i_{d2})}{dt} = \frac{1}{R}\left(\sqrt{\frac{2i_{d2}}{k_n}} - \sqrt{\frac{2i_{d1}}{k_n}}\right) + i_{d1} - i_{d2} \quad (15)$$

The normalized differential mode signal (common mode signal already given as $\frac{I_0}{2}$ before (8)), $\Delta = \frac{i_{d1}-i_{d2}}{I_0}$ is substituted into (15) to yield (16):

$$\frac{C_{gs}}{k_n\left(\sqrt{\frac{2i_d}{k_n}} - v_t\right)}\frac{d\Delta}{dt} = \frac{-1}{2Rk_n\left(\sqrt{\frac{2i_d}{k_n}} - v_t\right)}(\sqrt{1+\Delta} - \sqrt{1-\Delta}) + \Delta \quad (16)$$

By using Taylor expansion at $\Delta = 0$, $\sqrt{1+\Delta} - \sqrt{1-\Delta} = \Delta + \frac{1}{8}\Delta^3$. Then, (16) can be approximated to

$$RC_{gs}\frac{d\Delta}{dt} = \frac{1}{2}(Rg_m - 1)\Delta - \frac{1}{8}\Delta^3 \quad (17)$$

where we also set $k_n\left(\sqrt{\frac{2i_d}{k_n}} - v_t\right) = g_m =$ avg transconductance of $M_1$ and $M_2$.

Now let us turn familiar physical system, "ball on the hill"/spring/pendulum (2 level mechanical system).

For a damped spring system (where viscous damping force much stronger than the inertia term), having a nonlinear force, $F(x) = k_1 x + k_2 x^3$, the system is described as[9]:

$$\alpha \dot{x} = k_1 x + k_2 x^3 \quad (18)$$

α is friction coefficient, $k_1$, $k_2$ are the stiffness coefficients, while x is position.

Comparing (17), (18) they have similar forms.

However the spring system has familiar formula for the potential energy, given as the integral on the right hand side (RHS) of (18):

$$E = \int k_1 x + k_2 x^3 \, dx = \frac{1}{2}k_1 x^2 \frac{1}{4}k_2 x^4 + Const\_spring \quad (19)$$

Therefore we are led to integrate the RHS of (17), and interpret the result as the 'potential energy' of the Schmitt trigger, with the provision of 1) scaling of $C_{gs}$ and 2) in $I_0/g_m$ in the variable:

$$E = -\frac{1}{C_{gs}}\left(\frac{(Rg_m - 1)}{4}\left(\left(\frac{\Delta \cdot I_0}{g_m}\right)C_{gs}\right)^2 - \left(\frac{1}{32}\left(\frac{\Delta \cdot I_0}{g_m}\right)C_{gs}\right)^4\right) + Const \quad (20)$$

E has unit of Joule.

Moreover the potential energy of (19) (20) is an even function, with 1 maximum, and sandwiched between two minima, another characteristic of a 2 level system.

To visualize this better, instead of spring system, we use pendulum as shown in Figure 6 (with weight at the top, resting on a rod) i.e. an inverted pendulum. It is at the metastable state, with maximum energy, and any perturbation with push it to the stable state. Again the potential energy versus position is like (20) [11]. We now normalize (20) as:

$$normalized\ E = -\left(\frac{(Rg_m - 1)}{4}\Delta^2 - \frac{1}{32}\Delta^4\right) + Const \quad (21)$$

A numerical example of this normalized E is plotted as the solid line in Figure 7.

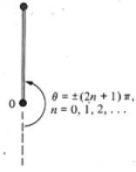

Figure 6      Inverted Pendulum system as 2 level system

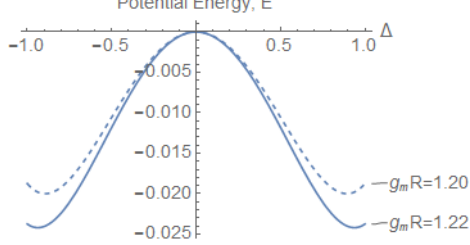

Figure 7 Potential Energy, E, normalized, vs Δ for parameter R=5kΩ, $Const = 0$. The dashed line is with the parameter $g_m R = 1.2$, and the solid line is with the parameter $g_m R = 1.22$

Now let us turn to the interpretation of (21), using this energy perspective. Here, Δ, identified earlier as the normalized differential current (see (16)), is now viewed as as a transformed coordinate [12] of the voltage $v_{gs1}$ and $v_{gs2}$, i.e. Δ is an algebraic function of $v_{gs1}, v_{gs2}$, in the form of $\Delta(v_{gs1}, v_{gs2}) \propto v_{gs1} - v_{gs2}$ (that is the reason for all the scaling done as explained prior to (21)). Since the electrostatic potential energy of a capacitor is $\frac{1}{2}CV_{gs}^2$, then this potential energy can also be seen in terms of Δ.

Now, first on the RHS of (16), the square root operates on this Δ, which reflects the nonlinearity of square law in the long channel I-V characteristic of $M_1$ and $M_2$. Secondly The "+Δ" and "-Δ" terms inside the square root on the RHS of (16) reflects the coupling interaction across $M_1$ and $M_2$. These translate, upon Taylor expansion, into the linear and cubic term i.e. Δ, $\Delta^3$ terms in (17), and subsequently, upon integration into the potential energy E, as $\Delta^2, \Delta^4$ terms in (21). In summary, E in (21) is rooted in $V_{gs1}$, $V_{gs2}$, (the co-ordinate in electrostatic potential energy), just like E in (19) is rooted in x, the co-ordinate in mechanical potential energy.

To define *Const*, we observe at metastable state, Δ=0, and so E=*Const*. The electrostatic potential energy, *Const* is from the common mode signal (explained before (8)) and so we can superpose the energy at both stable states and take the average. Thus, $Const=\left(\frac{Rg_m-1}{4} - \frac{1}{32}\right)$ when the stable states are Δ=1 and Δ=-1. The definition of *Const* is the sum potential energy of the parasitic capacitors (in transformed co-ordinate) at the metastable state. We then normalize[1] by setting *Const* to 0 and so E=0. Figure 7 shows that the potential energy of the Schmitt trigger is maximum (E=0) at metastable state. When the bistable system reach its stable states, the potential energy is at its minimum (E<0).

$\Delta E = E_{meta} - E_{stable}$ (the difference in potential energy at metastable state and stable state) is an important parameter that dictated the effectiveness of the system. This can be derived by differentiating (21) with respect to Δ, setting to 0, determine the resulting Δ, i.e. $\Delta|_{\frac{dE}{d\Delta}=0}$, substituting back in (20), giving ΔE as

$$\Delta E = \frac{1}{2}(g_m R - 1)^2 \quad (22)$$

It shows that when $g_m R \to 1$, ΔE goes to zero and the system behaves less and less like a 2 level system, and hence will not work as an effective Schmitt trigger. This is evident in Figure 7, when $g_m R$ is lowered from 1.22 (solid line) to 1.2 (dash line), ΔE is smaller.

## V. CONCLUSION

A novel way of analyzing Schmitt trigger as 2 level (metastable/stable) energy system is presented. The energy of the electronic 2 level system (Schmitt trigger) is identified. The potential energy of the Schmitt trigger reflects the electrostatic potential energy. The simulation verifies the potential energy of the system as a 2 level bistable (metastable/stable) system. New design insight, using this physics based model, is presented.

---

[1] This is like the potential energy of a test charge q for a given charge, Q: When q is far away (∞ distance) from Q, the potential energy is zero and is at its maximum. When charge q closes in, the electrostatic attraction causes the potential energy to be negative.